\documentclass[runningheads]{llncs}
\usepackage[T1]{fontenc}
\usepackage{cite}
\usepackage{amsmath} 
\usepackage{hyperref}
\usepackage[a4paper, margin=1.4in]{geometry}

\usepackage{graphicx}
\usepackage{color}

\begin{document}
%
\title{Comparative Analysis of nnUNet and MedNeXt for Head and Neck Tumor Segmentation in MRI-guided Radiotherapy}
\titlerunning{Comparison of Automatic Segmentation Frameworks}
%
\author{Nikoo Moradi \inst{1,2} 
\and
André Ferreira \inst{2,3,4,5,6} 
\and 
Behrus Puladi\inst{5,6}
\and 
Jens Kleesiek\inst{2,7,8,9} 
\and 
Emad Fatemizadeh\inst{1} 
\and 
Gijs Luijten \inst{2, 10, 11}
\and
Victor Alves \inst{3}
\and
Jan Egger\inst{2,4,7,10,11}}
\authorrunning{N. Moradi et al.}
%
\institute{Department of Electrical Engineering, Sharif University of Technology, Tehran, Iran
\and
Institute for AI in Medicine (IKIM), University Medicine Essen, Girardetstraße 2, Essen, 45131, Germany 
\and
Center Algoritmi / LASI, University of Minho, Braga, 4710-057,  Portugal
\and
Computer Algorithms for Medicine Laboratory, Graz, Austria
\and 
Department of Oral and Maxillofacial Surgery, University Hospital RWTH Aachen, Aachen,  Germany
\and 
Institute of Medical Informatics, University Hospital RWTH Aachen, Aachen, Germany
\and
Cancer Research Center Cologne Essen (CCCE), University Medicine Essen (AöR), Essen, Germany
\and
German Cancer Consortium (DKTK), Partner Site Essen, Essen, Germany
\and
Department of Physics, TU Dortmund University, Dortmund, Germany
\and
Center for Virtual and Extended Reality in Medicine (ZvRM), University Medicine Essen, Essen, Germany
\and
Institute of Computer Graphics and Vision (ICG), Graz University of Technology, Inffeldgasse 16/II, 8010, Graz, Austria
\\
\email{nikoomoradi81@gmail.com}}
\maketitle              
\begin{abstract}
Radiation therapy (RT) is essential in treating head and neck cancer (HNC), with magnetic resonance imaging(MRI)-guided RT offering superior soft tissue contrast and functional imaging. However, manual tumor segmentation is time-consuming and complex, and therfore remains a challenge. In this study, we present our solution as team TUMOR to the HNTS-MRG24 MICCAI Challenge which is focused on automated segmentation of primary gross tumor volumes (GTVp) and metastatic lymph node gross tumor volume (GTVn) in pre-RT and mid-RT MRI images.
We utilized the HNTS-MRG2024 dataset, which consists of 150 MRI scans from patients diagnosed with HNC, including original and registered pre-RT and mid-RT T2-weighted images with corresponding segmentation masks for GTVp and GTVn. We employed two state-of-the-art models in deep learning, nnUNet and MedNeXt. For Task 1, we pretrained models on pre-RT registered and mid-RT images, followed by fine-tuning on original pre-RT images. For Task 2, we combined registered pre-RT images, registered pre-RT segmentation masks, and mid-RT data as a multi-channel input for training. Our solution for \textbf{Task 1} achieved 1st place in the final test phase with an aggregated Dice Similarity Coefficient of \textbf{0.8254}, and our solution for \textbf{Task 2} ranked 8th with a score of \textbf{0.7005}. The proposed solution is publicly available at 
\href{https://github.com/NikooMoradi/HNTSMRG24_team_TUMOR}{Github Repository}.

\end{abstract}
\section{Introduction}

Radiation therapy (RT) is a fundamental treatment modality for various malignancies, with head and neck cancer (HNC) being a primary beneficiary. Traditional RT planning has largely relied on computed tomography (CT) imaging. However, recent advancements have driven significant interest in magnetic resonance imaging (MRI)-guided RT. MRI offers superior soft tissue contrast compared to CT and enables functional imaging through multiparametric sequences, such as diffusion-weighted imaging. Additionally, MRI-guided RT facilitates daily adaptive treatment using MRI-Linac devices, optimizing tumor destruction while minimizing adverse effects. These advantages suggest that MRI-guided adaptive RT has the potential to revolutionize clinical practice for HNC. \cite{Kiser2019DataFlood,Pollard2017MRGuided}

Despite these benefits, MRI-guided RT planning generates extensive data, making manual tumor segmentation by physicians—the current clinical standard—a time-consuming and impractical process. This challenge is intensified by the complex anatomy of head and neck (H\&N) tumors, which are notoriously difficult to delineate accurately. As a result, there is a growing interest in leveraging artificial intelligence (AI) to automate and improve the segmentation process.

Deep learning (DL), a subset of AI, has shown remarkable success in medical image segmentation, particularly in challenging domains like HNC. Various public challenges, such as the HECKTOR \cite{Andrearczyk2022} and SegRap \cite{SegRap2023} challenges, have driven advancements in this field by providing datasets and benchmarks for AI model development. However, no large-scale, publicly available datasets for MRI-guided RT in HNC exist, highlighting the need for community-driven efforts to develop AI tools for clinical translation.

The HNTS-MRG24 challenge \footnote{\href{https://hntsmrg24.grand-challenge.org}{https://hntsmrg24.grand-challenge.org}} addresses this gap by focusing on the segmentation of H\&N tumors in MRI-guided adaptive RT. The challenge is divided into two tasks: 
\subsubsection{Task 1:} Segmentation of primary gross tumor volume (GTVp) and metastatic node gross tumor volume (GTVn) on pre-RT MRI images.
\subsubsection{Task 2:} Extends this to mid-RT MRI images. In this task mid-RT image, pre-RT image with segmentation, and registered pre-RT image with registered segmentation are all available and can be used as input.

A unique aspect of this challenge is its exploration of whether incorporating prior time point data (pre-RT and mid-RT) into segmentation algorithms can enhance performance in RT applications.

Given the potential of AI to streamline and enhance MRI-guided RT planning, the development of robust, automated segmentation algorithms could significantly impact clinical workflows, reducing the burden on clinicians and improving patient outcomes. 

This paper presents our approach to the HNTS-MRG24 challenge, utilizing state-of-the-art DL models, i.e, nnUNet \cite{isensee2021nnunet} and MedNeXt \cite{Roy2023MedNeXt,isensee2020nnunet}, and ensemble techniques to achieve accurate and reliable segmentation of H\&N tumors.

\subsection*{State of the art}

Recent advancements in AI-based segmentation of HNC have demonstrated significant potential, particularly in the context of RT planning. Various studies have utilized DL models to automate and improve the segmentation of tumors and organs at risk, addressing the challenges posed by the complex anatomy of the H\&N region.

Li et al. (2020) \cite{Li2020GenericEnsemble} proposed a semi-supervised framework for medical image segmentation using deep convolutional neural networks. Their method combines labeled and unlabeled data to improve segmentation accuracy by generating pseudo labels and iteratively refining the model. The framework incorporates ensemble learning to reduce errors from poor-quality pseudo labels. The approach was evaluated on the ISIC 2018 dataset for skin lesion segmentation \cite{Codella2019ISICChallenge,Tschandl2018HAM10000}, and it demonstrated superior performance compared to fully supervised models and earlier semi-supervised methods.

Astaraki et al. (2023) \cite{Astaraki2023SegRap} focused on nasopharyngeal carcinoma, a subset of HNC, in the SegRap 2023 challenge. They developed a fully automated segmentation framework using a standard 3D U-Net model, which was effective in segmenting both GTVs and organs at risk from CT images. Their approach achieved first place in the second task of SegRap 2023 challenge, underscoring the robustness of the U-Net architecture for this task.

Myronenko et al. (2022)\cite{Myronenko2022HECKTOR} presented a fully automated solution for H\&N tumor segmentation using positron emission tomography (PET)/CT images in the HECKTOR 2022 \cite{Myronenko2022HECKTOR} challenge. They employed the SegResNet architecture from MONAI, a semantic segmentation network optimized for 3D medical imaging. Their approach included 5-fold cross-validation, image normalization, and model ensembling, which helped achieve first place in the challenge.

These studies collectively highlight the ongoing efforts to refine DL techniques for HNC segmentation, with a particular focus on improving accuracy, handling data imbalance, and integrating multi-modal imaging data. The continued development of AI-driven segmentation tools holds promise for enhancing the precision and efficiency of RT planning, ultimately improving patient outcomes.

The remainder of this paper is structured as follows: Section 2 covers the materials and methods used in our approach, Section 3 presents the results and evaluation of our models, Section 4 discusses the findings, and Section 5 concludes with a summary and future directions.

\section{Materials and Methods}

\subsection{Dataset}

\subsubsection{HNTS-MRG24 Dataset:}
For this study, we utilized the dataset provided by the HNTS-MRG24 challenge, which focuses on the segmentation of H\&N tumors for MRI-guided adaptive RT. The dataset comprises MRI images from 150 patients diagnosed with HNC, collected at The University of Texas MD Anderson Cancer Center. It includes both pre-RT and mid-RT T2-weighted (T2w) MRI scans, with corresponding segmentation masks for GTVp and GTVn.

An important aspect of the dataset is the pre-registration of pre-RT images to mid-RT images, which was performed by the challenge organizers. This registration process was designed to align the images spatially, facilitating more accurate comparison and analysis between the different time points. Detailed parameters for this registration process can be found in the challenge's official GitHub repository \footnote{\href{https://github.com/kwahid/HNTSMRG\_2024}{https://github.com/kwahid/HNTSMRG\_2024}}.

For Task 1, we utilized the mid-RT and pre-RT registered images to pretrain our models, followed by fine-tuning on the original pre-RT images. During this phase, all 150 cases were included, and no cases were discarded.

\begin{figure}
    \centering
    \includegraphics[width=0.7\linewidth]{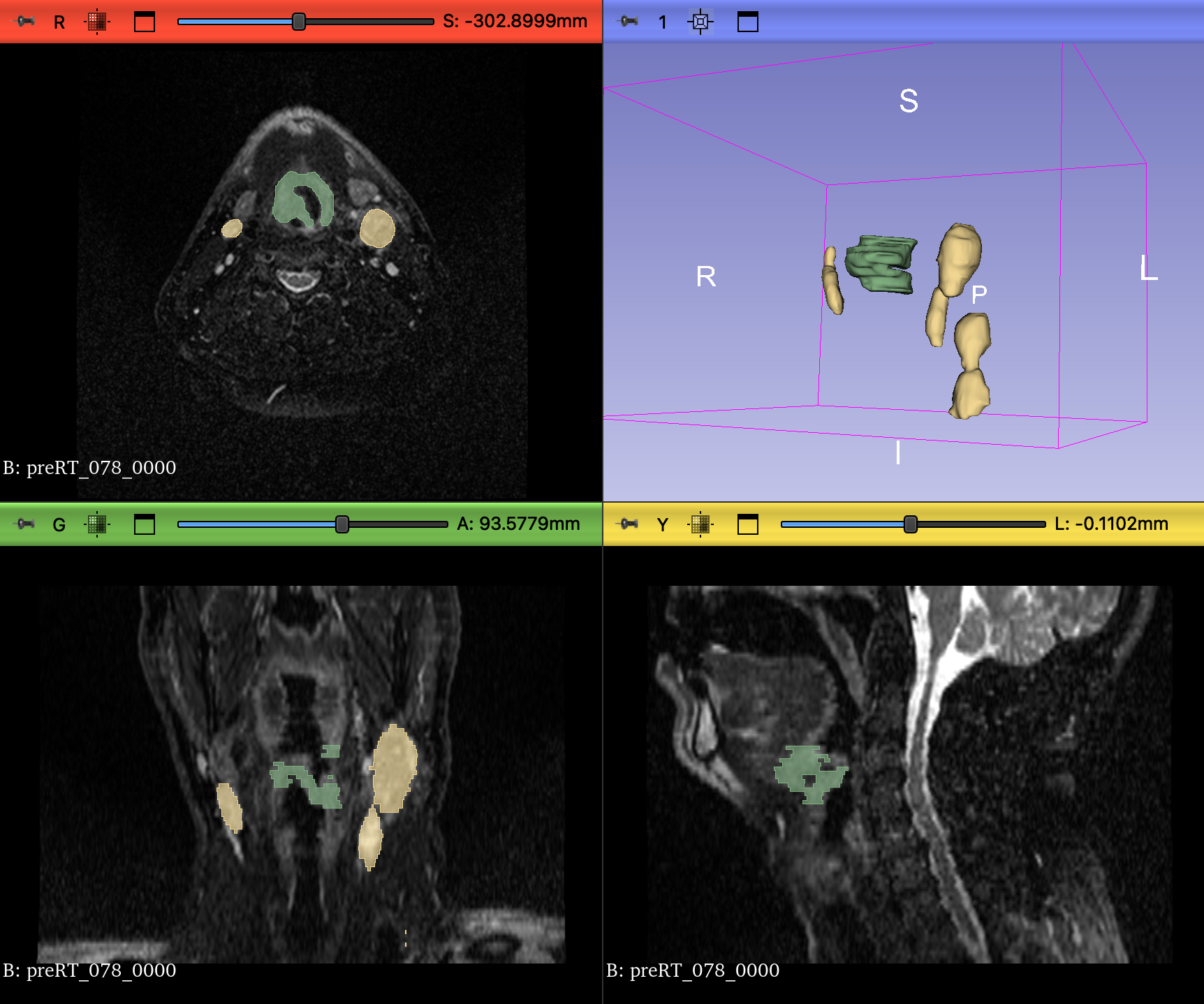}
    \caption{A sample pre-RT image (Case 78) with its corresponding segmentation. The green label represents GTVp (label=1), while the yellow label represents GTVn (label=2). The images show axial, coronal, and sagittal views, along with a 3D rendering of the segmented tumors.}
    \label{fig:fig1}
\end{figure}

For Task 2, the input consisted of a multi-channel format combining mid-RT, pre-RT registered images, and the corresponding pre-RT registered segmentation. All 150 cases were used to train nnUNet, but we encountered issues while training MedNeXt. To resolve these issues, cases with zero ground truth for either label 1, label 2, or both were discarded, leaving 115 samples after removing 35 cases.
An example of a pre-RT image with its segmentation is shown in Figure \ref{fig:fig1}. All the visualizations in this paper were created using 3D Slicer \cite{1398617}.

\subsubsection{BraTS24 Meningioma Radiotherapy Dataset:}

In addition to the challenge data, external public datasets were permitted. However, finding datasets that matched the challenge criteria (T2w MRI, at least two segmentation labels, H\&N region) proved challenging. We experimented with the BraTS 2024 Meningioma RT Segmentation Challenge dataset \cite{LaBella2024BraTS}, which contains 500 samples and focuses on the segmentation of GTV for meningiomas in brain MRI scans. This dataset uses 3D postcontrast T1w images and preserves extracranial structures through defacing techniques.

We used all 500 cases of the BraTS dataset specifically for Task 1 of the challenge and applied it as pretraining data for our models.

\subsection{Networks}
\label{subsec:Networks}

The effectiveness of two state-of-the-art models in DL, nnUNet and MedNeXt, was tested in both tasks of the HNTS-MRG24 challenge. Their performance was compared across different architectures and training strategies. A detailed description of the architectures used for each task will follow.

\subsubsection{nnUNet:}
nnUNet is an automated DL framework that self-configures based on the properties of the input data. It has proven effective across a wide range of medical image segmentation tasks due to its adaptability and robust performance \cite{isensee2021nnunet}. Specifically, we employed the following architectures, each for 5 folds:
\begin{itemize}
    \item 3D Full Resolution (FullRes) U-Net with the default planner
    \item 3D U-Net Cascade with the default planner
    \item 3D FullRes U-Net with Large Residual Encoder (ResEnc) Presets (nnUNetPlannerResEncLPlans)

\end{itemize}

For further information about nnUNet different configurations and planners, please refer to its documentation \footnote{\href{https://github.com/MIC-DKFZ/nnUNet/tree/master/documentation}{https://github.com/MIC-DKFZ/nnUNet/tree/master/documentation}}.

\subsubsection{MedNeXt:}
MedNeXt is a DL architecture tailored for medical image analysis, particularly effective in handling varying image modalities \cite{Roy2023MedNeXt,isensee2020nnunet}. We utilized the following architectures, each for 5 folds:
\begin{itemize}
    \item Small model with 3x3x3 kernel size
    \item Small model with 5x5x5 kernel size
    \item Large model with 3x3x3 kernel size
    \item Large model with 5x5x5 kernel size
\end{itemize}

 When referring to small and large MedNeXt models, the terms relate to the compound scaling of the model. This refers to the simultaneous scaling of depth (number of layers), width (number of channels), and receptive field (kernel size).
The small functional design (MedNeXt-S) utilizes 32 channels, an expansion ratio of 2, and a block count of 2.
The largest architecture (MedNeXt-L) consists of 62 MedNeXt blocks and uses high values of both expansion ratio and block count \cite{Roy2023MedNeXt}.

The decision to explore these configurations (kernel sizes 3x3x3 and 5x5x5 for small and large models) is supported by findings in the MedNeXt study \cite{Roy2023MedNeXt}. Smaller kernels, such as 3x3x3, provide a robust baseline with balanced computational efficiency and performance. Larger kernels, such as 5x5x5, leverage the ability of MedNeXt to learn long-range spatial dependencies, which are particularly useful for medical images with complex anatomical structures.

The MedNeXt study also demonstrates that MedNeXt-L outperforms or is competitive with smaller variants across tasks involving heterogeneous datasets (brain and kidney tumors, organs), varying modalities (CT, MRI), and diverse training set sizes. On the other hand MedNeXt-S can be more computationally efficient and data-efficient, which are important considerations in medical image segmentation where computational resources may be limited and datasets are often small \cite{Roy2023MedNeXt}.

For further information about different configurations of MedNeXt, please visit its documentation \footnote{\href{https://github.com/MIC-DKFZ/MedNeXt}{https://github.com/MIC-DKFZ/MedNeXt}}.

\subsection{Ensemble strategy}
To enhance the performance of our segmentation models, we implemented a multi-level ensemble strategy. First, we applied the default ensembling methods for each model framework. For nnUNet models, we used \texttt{nnUNetv2\_ensemble} and for MedNeXt models, we used \texttt{MedNeXtv1\_ensemble} to combine the predictions from different architectures of each model. 
\texttt{nnUNetv2\_ensemble} and \texttt{MedNeXtv1\_ensemble} are the commands provided by the frameworks, which are average ensemble.

After obtaining the ensembled predictions from nnUNet and MedNeXt separately, average ensemble method was applied on these outputs to produce the final segmentation result. However, due to differences in the size of the probability maps generated by MedNeXt and nnUNet, additional preprocessing was required. Specifically, we padded the probability maps from MedNeXt to match the size of the original compressed nifti images. Once the probability maps were aligned, we computed average and converted the averaged probability map into a segmentation image.

\subsection{Methodology}
The training and ensemble strategies, as well as the incorporation of external data, are described below.

\subsubsection{Task 1:} 
For Task 1, different configurations of nnUNet and MedNeXt (as described in section \ref{subsec:Networks}) were pre-trained on mid-RT and pre-RT registered images as individual inputs, and then fine-tuned on the original pre-RT images. After training all the models, ensembling strategies were applied to improve performance. Initially, every possible combination of nnUNet models (FullRes, Cascade, and ResEnc) was aggregated. Following this, the outputs of MedNeXt were combined with the best-performing combination of nnUNet models to evaluate possible improvements.

To explore the effectiveness of the BraTS dataset for the HNC segmentation task, nnUNet FullRes was first pretrained on the BraTS dataset (500 samples) and then fine-tuned on the original pre-RT images. For this experiment, to handle the transition from the BraTS dataset (a single-channel output) to the HNTS-MRG24 dataset requiring segmentation of two channels (GTVp and GTVn), two labels were assigned instead of one during the pretraining phase. This adjustment ensured that the output layer of the pretrained model had two channels, and it is compatible with fine-tuning on the HNTS-MRG24 dataset without further modification to the output layer.

Additionally, nnUNet FullRes was pretrained on a combined dataset of BraTS and mid-RT and pre-RT registered images (800 samples) to assess the benefit of combining external data with challenge-specific data. Following this, fine-tuning was performed on the original pre-RT images.

\subsubsection{Task 2:} To evaluate the potential value of pre-RT images and their segmentation masks for mid-RT segmentation, nnUNet FullRes and MedNeXt small model with kernel size 3 were trained on four different datasets:
\begin{itemize}
\label{task2:datasets}
    \item Mid-RT images only, referred to as Dataset 504.
    \item Mid-RT and registered pre-RT images as a multi-channel input, referred to as Dataset 505.
    \item Mid-RT images, registered pre-RT images, and pre-RT segmentation masks as a multi-channel input, referred to as Dataset 506.
    \item Mid-RT and registered pre-RT segmentation masks as a multi-channel input, referred to as Dataset 507.
\end{itemize}

After identifying the best dataset, all model configurations mentioned in section \ref{subsec:Networks} were trained on it. The same ensembling strategy as in Task 1 was applied, first aggregating combinations of nnUNet models (FullRes, Cascade, ResEnc) and then combining the best-performing combination of nnUNet models outputs with MedNeXt predictions to evaluate possible improvements.

\subsection{Evaluation}
The validation set was evaluated using the Aggregated Dice Similarity Coefficient ($\text{DSC}_{\text{agg}}$) \cite{9871907} , same as the challenge evaluation standards.
\begin{equation}
\text{DSC}_{\text{agg}} = \frac{2 \sum_{i} |A_i \cap B_i|}{\sum_{i} \left( |A_i| + |B_i| \right) }
\label{eq:1}
\end{equation}

In this context, $A_i$ and $B_i$ represent the ground truth and predicted segmentations for image $i$, respectively, where $i$ ranges across the entire test set.

Additionally, DSC was calculated for each label (GTVp, GTVn) on a per-sample basis, for each model \cite{ed278621-dc3e-343f-ae66-540d8990b60d}. 
\begin{equation}
\text{DSC}_{\text{i}} = \frac{2 |A_i \cap B_i|}{|A_i| + |B_i|}
\label{eq:2}
\end{equation}

In this context, $A_i$ represents the ground truth and $B_i$ represents the predicted segmentations for image $i$. 

For cases with zero ground truth (\( |A_i| = 0 \)), the predictions were checked to determine whether the model produced a true empty segmentation (no tumor predicted for a sample with zero ground truth). In such cases, the DSC was assigned a value of 1. Conversely, if the model produced a non-empty segmentation for a sample with zero ground truth, the DSC was assigned a value of 0. After addressing these scenarios, the mean and standard deviation (STD) of the DSC were calculated across all samples to further assess model performance.

\section{Results}
The experiments were conducted using the cluster node of the Institute for Artificial Intelligence in Medicine (IKIM) in Essen, Germany. The node has 6 NVIDIA RTX 6000, 48 GB of VRAM, 1024 GB of RAM, and AMD EPYC 7402 24-Core Processor.
The software environment included Python 3.9.19, PyTorch version 2.3.1+cu121, nnUNet version 2.5, and MedNeXt version 1.7.0.

\subsection*{Task 1:} The performance of different configurations of the nnUNet and MedNeXt models was evaluated using $\text{DSC}_{\text{agg}}$. While we intended to train all configurations of both models, we encountered issues with certain MedNeXt configurations. The small model with kernel size 3 was the only one that trained successfully and other configurations kept collapsing after a few epochs. 
On the other hand, nnUNet was successfully trained in all three configurations: FullRes, ResEnc, and Cascade.

For nnUNet, we applied its default ensembling strategy (\texttt{nnUNetv2\_ensemble}) to combine predictions from the FullRes, ResEnc, and Cascade configurations. Every possible combination of these models was ensembled. We also attempted an additional step of averaging the best predictions from nnUNet (ensemble of Cascade and ResEnd) and MedNeXt. An overview of these results is given in Table \ref{tab1}.

In addition to $\text{DSC}_{\text{agg}}$, mean DSC and STD were also calculated the for each predicted label (GTVp, GTVn) across all samples for each model configurations to further investigate model stability and robustness under different conditions. These results are shown in Table \ref{tab1.1}.

\begin{figure}
    \centering
    \includegraphics[width=1\linewidth]{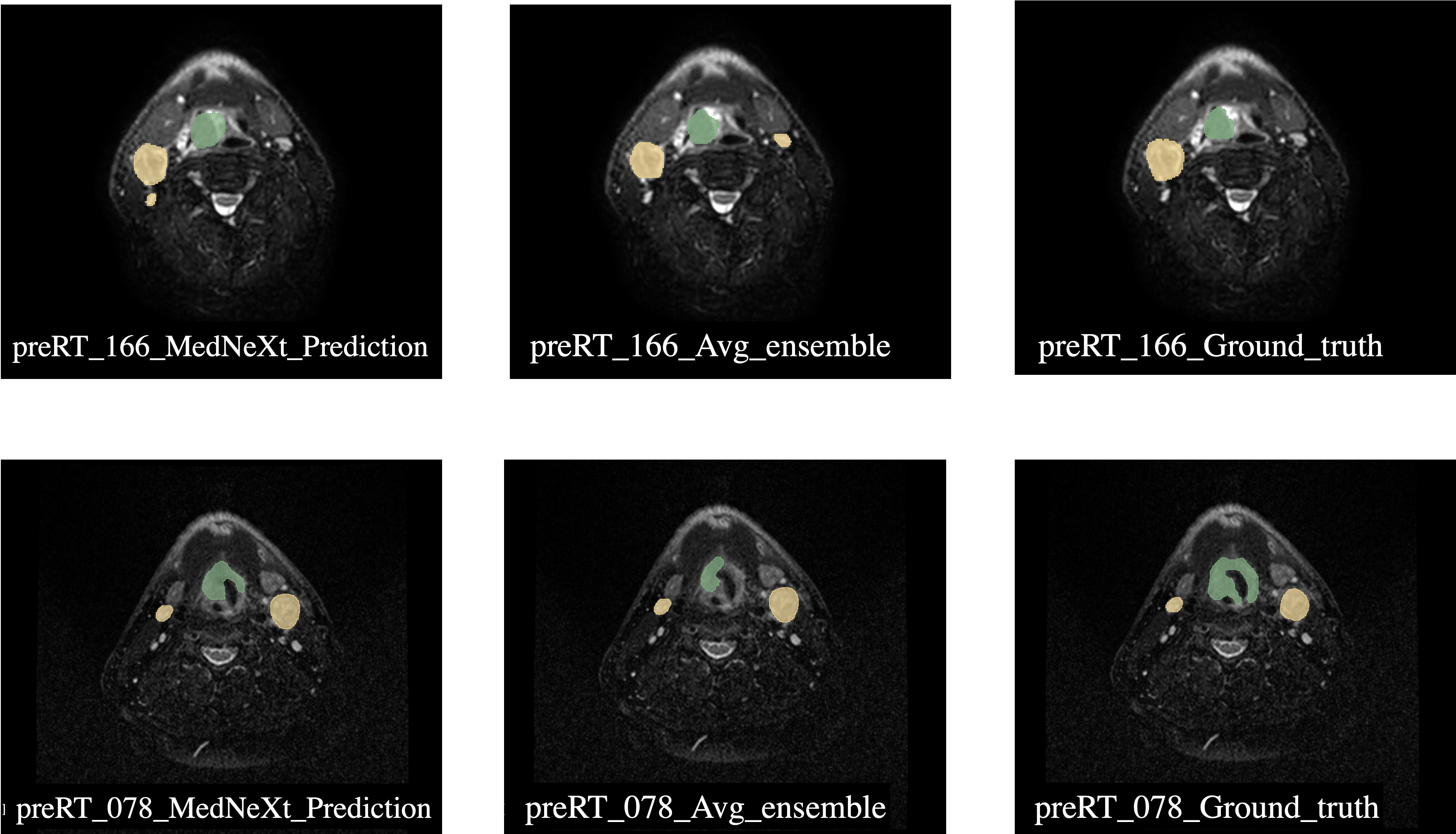}
    \caption{Comparison of predicted segmentations of two pre-RT samples (Case 78 and 166) for Task 1. The left image shows the prediction from MedNeXt with the best DSC${\text{agg}}$, the middle image shows the prediction from the average ensemble of nnUNet and MedNeXt, which had the lowest DSC${\text{agg}}$, and the right image shows the ground truth segmentation. The green label represents GTVp (label=1), and the yellow label represents GTVn (label=2).}
    \label{fig:fig2}
\end{figure}

Figure \ref{fig:fig2} shows a comparison between the best-performing MedNeXt prediction, the worst-performing average ensemble of nnUNet and MedNeXt, and the ground truth segmentation.

\begin{table}
\centering
\caption{$\text{DSC}_{\text{agg}}$ for each model configuration for Task 1}\label{tab1}
\begin{tabular}{|l|c|c|c|}
\hline
\textbf{Model} & \textbf{ GTVp $\text{DSC}_{\text{agg}}$ } & \textbf{ GTVn $\text{DSC}_{\text{agg}}$ } & \textbf{ Mean } \\
\hline
nnUNet FullRes & 0.7772 & 0.8517 & 0.8144 \\
nnUNet ResEnc & 0.7873 & 0.8586 & 0.8230 \\
nnUNet Cascade & 0.7847 & 0.8550 & 0.8198 \\
nnUNet Cascade + FullRes & 0.7846 & 0.8573 & 0.8210 \\
nnUNet Cascade + ResEnc & 0.7919 & 0.8633 & 0.8276 \\
nnUNet FullRes + ResEnc & 0.7896 & 0.8601 & 0.8249 \\
nnUNet All Ensembled & 0.7889 & 0.8618 & 0.8254 \\
\hline
MedNeXt Small (Kernel 3) & {\bfseries0.8066} & 
{\bfseries0.8710} & {\bfseries0.8388}\\
\hline
nnUNet + MedNeXt (Average) & 0.7931 & 0.8166 & 0.8049 \\
\hline
\end{tabular}
\end{table}

\begin{table}
\centering
\caption{Mean DSC $\pm$ STD over all cases for each model configuration for Task 1}\label{tab1.1}
\begin{tabular}{|l|c|c|}
\hline
\textbf{Model} & \textbf{ GTVp DSC } & \textbf{ GTVn DSC } \\
\hline
nnUNet FullRes & 0.6101 $\pm$ 0.3453 & 0.7488 $\pm$ 0.2761 \\
nnUNet ResEnc & 0.6066 $\pm$ 0.3509 & 0.7619 $\pm$ 0.2750 \\
nnUNet Cascade & 0.6272 $\pm$ 0.3331 & 0.7431 $\pm$ 0.2772 \\
nnUNet Cascade + FullRes & 0.6247 $\pm$ 0.3392 & 0.7502 $\pm$ 0.2762 \\
nnUNet Cascade + ResEnc & 0.6359 $\pm$ 0.3402 & 0.7701 $\pm$ 0.2700 \\
nnUNet FullRes + ResEnc & 0.6283 $\pm$ 0.3483 & 0.7765 $\pm$ 0.2611 \\
nnUNet All & 0.6265 $\pm$ 0.3405 & 0.7682 $\pm$ 0.2630 \\
\hline
MedNeXt Small (Kernel 3) & \textbf{0.6940 $\pm$ 0.2982} & \textbf{0.8010 $\pm$ 0.2341} \\
\hline
nnUNet + MedNeXt (Average) & 0.6636 $\pm$ 0.3147 & 0.7654 $\pm$ 0.2520 \\
\hline
\end{tabular}
\end{table}

Additionally inference of pretrained models compared to fine-tuned models was tested, see Table \ref{tab1} and \ref{tab2}. Interestingly, these pretrained models performed better on the original pre-RT images compared to the models that were fine-tuned on the original pre-RT images. 

\begin{table}
\centering
\caption{$\text{DSC}_{\text{agg}}$ for Pretrained Models on Pre-RT Registered and Mid-RT Images}\label{tab2}
\begin{tabular}{|l|c|c|c|}
\hline
\textbf{Model } & \textbf{ GTVp $\text{DSC}_{\text{agg}}$ } & \textbf{ GTVn $\text{DSC}_{\text{agg}}$ } & \textbf{ Mean } \\
\hline
Pretrained nnUNet FullRes & 0.8545 & 0.8930 & 0.8737 \\
Pretrained nnUNet Cascade & 0.8701 & 0.8942 & 0.8822 \\
Pretrained nnUNet ResEnc & \textbf{0.8936} & \textbf{0.9197} & \textbf{0.9066} \\
\hline
Pretrained MedNeXt Small (Kernel 3) & 0.8748 & 0.8994 & 0.8871 \\
\hline
\end{tabular}
\end{table}

We experimented with using the BraTS dataset as external data. It was used in the pretraining step, either alone or combined with mid-RT and pre-RT registered images, and then fine-tuned on the original pre-RT images. The results of these experiments are presented in Table \ref{tab3}.


\begin{table}[h!]
\centering
\caption{$\text{DSC}_{\text{agg}}$ for nnUNet FullRes Pretrained on Different Datasets and Fine-Tuned on Original Pre-RT for Task 1}\label{tab3}
\begin{tabular}{|l|c|c|c|c|}
\hline
\textbf{Pretrain on} & \textbf{Fine-tune on} & \textbf{GTVp $\text{DSC}_{\text{agg}}$} & \textbf{GTVn $\text{DSC}_{\text{agg}}$} & \textbf{Mean} \\
\hline
BraTS & \_  & 0.0904 & 0.0000 & 0.0452 \\
BraTS & Original pre-RT & 0.6871 & 0.7605 & 0.7238 \\
\hline
BraTS+mid-RT+Reg pre-RT & \_ & 0.8347 & 0.8740 & 0.8544 \\
BraTS+mid-RT+Reg pre-RT & Original pre-RT & 0.7611 & 0.8473 & 0.8042 \\
\hline
mid-RT+Reg pre-RT & \_ & \textbf{0.8545} & \textbf{0.8930} & \textbf{0.8737} \\
mid-RT+Reg pre-RT & Original pre-RT & 0.7772 & 0.8517 & 0.8144 \\
\hline
\end{tabular}
\end{table}

For Task 1, the final submission used the MedNeXt small model with kernel size 3. It achieved a $\text{DSC}_{\text{agg}}$ of \textbf{0.8728} for GTVn and \textbf{0.7780} for GTVp, with an overall mean $\text{DSC}_{\text{agg}}$ of \textbf{0.8254} in the final test phase on the 50 test patients.

\subsection*{Task 2:}
For this task, nnUNet FullRes and MedNeXt small model with kernel size 3 were trained on four datasets with different combinations of mid-RT images, registered pre-RT images, and their segmentation masks (see \ref{task2:datasets}).

nnUNet FullRes was successfully trained for all experiments. However, while training MedNeXt small with kernel size 3 on Dataset 506, the model collapsed after a few hundred epochs for some folds. Table \ref{tab4} outlines the number of epochs each fold was trained for. Despite the incomplete training, we proceeded with inference for MedNeXt small model with kernel size 3 on Dataset 506.

\begin{table}[h!]
\centering
\caption{Training Epochs for MedNeXt (Small with Kernel size 3) on Dataset 506 Across Folds}\label{tab4}
\begin{tabular}{|c|c|c|c|c|c|}
\hline
 & \textbf{ fold 0 } & \textbf{ fold 1 } & \textbf{ fold 2 } & \textbf{ fold 3 } & \textbf{ fold 4 } \\
\hline
 Num of epochs  & 344 & 1000  & 1000 & 626 & 672 \\
\hline
\end{tabular}
\end{table}

To address the training issues with MedNeXt on dataset 506, we discarded 35 samples where either label 1, label 2, or both were zero in the segmentation mask of the registered pre-RT. This resulted in 115 samples (dataset 516), which were then used to train MedNeXt.
 The results are presented in Table \ref{tab5}.

\begin{table}[h!]
\centering
\caption{$\text{DSC}_{\text{agg}}$ for nnUNet FullRes and MedNeXt Small with Kernel size 3 Trained on Different Datasets for Task 2}\label{tab5}
\begin{tabular}{|l|c|c|c|c|}
\hline
\textbf{Model} & \textbf{Trained on} & \textbf{GTVp $\text{DSC}_{\text{agg}}$} & \textbf{GTVn $\text{DSC}_{\text{agg}}$} & \textbf{Mean} \\
\hline
nnUNet FullRes & Dataset 504  & 0.5000 & 0.8085 & 0.6542 \\
    nnUNet FullRes & Dataset 505 & 0.4608 & 0.8025 & 0.6316 \\
nnUNet FullRes & Dataset 506  & \textbf{0.6100} & \textbf{0.8508} & \textbf{0.7304} \\
nnUNet FullRes & Dataset 507  & 0.6007 & 0.8513 & 0.7260 \\
\hline
MedNeXt  Small (Kernel 3) & Dataset 504 & 0.5676 & 0.8162 & 0.6919 \\
MedNeXt  Small (Kernel 3) & Dataset 505 & 0.5678 & 0.8091 & 0.6884 \\
MedNeXt  Small (Kernel 3) & Dataset 506 & 0.6099 & 0.8306 & 0.7202 \\
MedNeXt  Small (Kernel 3) & Dataset 507 & \textbf{0.6231} & \textbf{0.8446} & \textbf{0.7339} \\
MedNeXt  Small (Kernel 3) & Dataset 516 & 0.5904 & 0.8275 & 0.7089 \\

\hline
\end{tabular}
\end{table}

After experimenting on different datasets and finding the best one, we used the same architectures for Task 2 as in Task 1 to compare different models and configurations and ensemble strategies. All configurations of nnUnet trained successfully. Other MedNeXt architectures (small model with kernel size 5, large model with kernel size 3, and large model with kernel size 5) collapsed after only a few epochs.
As with Task 1, we ensembled all possible combinations of nnUNet predictions using default nnUNet ensembling (\texttt{nnUNetv2\_ensemble}), and then ensembled the best nnUNet predictions (ensemble of Cascade and FullRes) with MedNeXt predictions (from dataset 506, despite incomplete training on some folds) using the average ensembling method. An overview is given in Table \ref{tab6}.

In addition to $\text{DSC}_{\text{agg}}$, we also calculated the mean DSC and STD for each predicted label (GTVp, GTVn) across all samples for each model configurations to further evaluate model robustness and performance consistency under varying conditions. These results are provided in Table \ref{tab8}.

\begin{figure}
    \centering
    \includegraphics[width=1\linewidth]{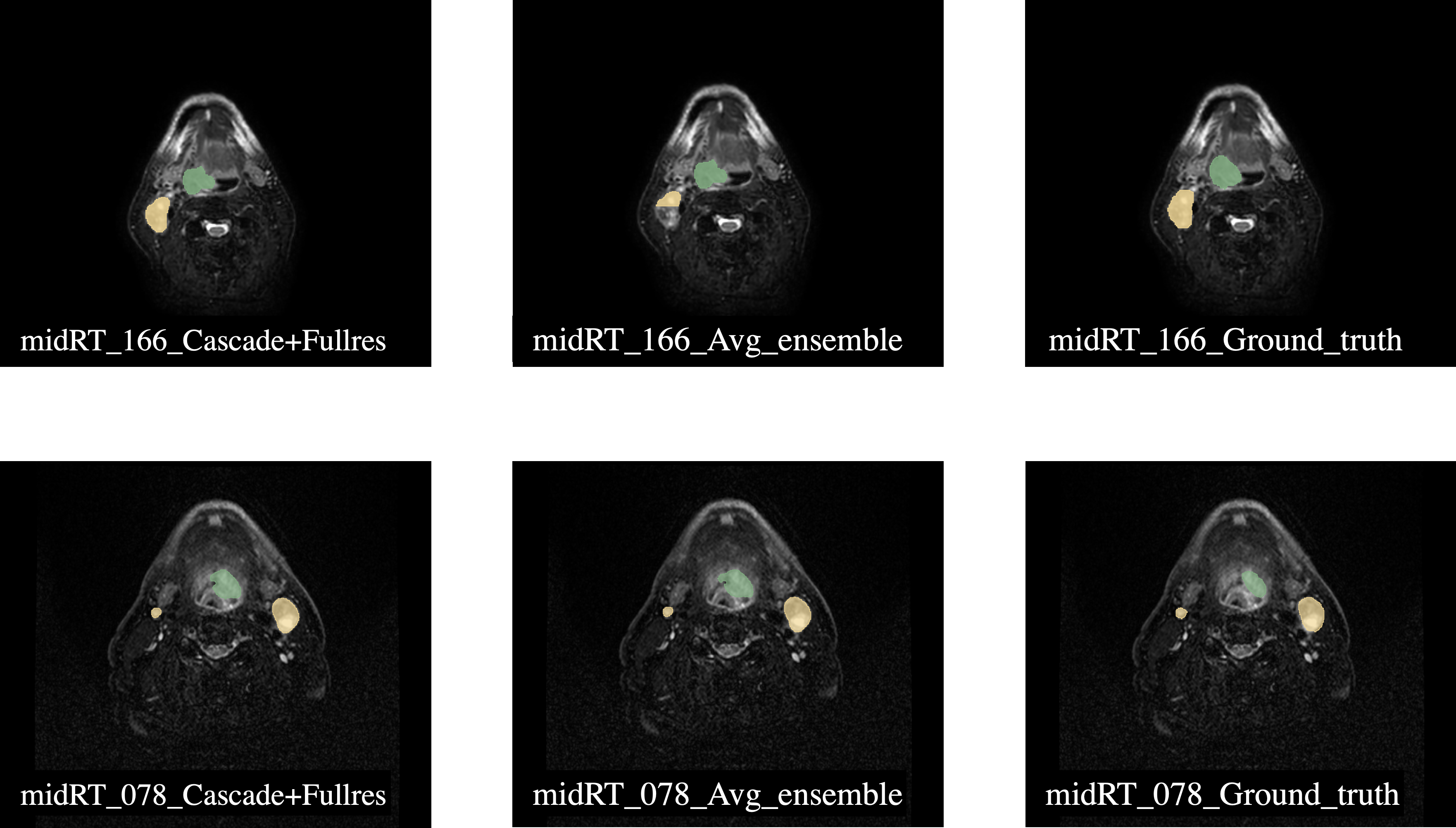}
    \caption{Comparison of segmentation predictions of sample mid-RT (Case 78 and 166) for Task 2. The left image shows the prediction from ensemble of nnUNet Cascade and FullRes with the best DSCagg, the middle image shows the prediction from the average ensemble of nnUNet and MedNeXt, which had the lowest DSCagg, and the right image shows the ground truth segmentation. The green label represents GTVp (label=1), and the yellow label represents GTVn (label=2).}
    \label{fig:fig3}
\end{figure}

Figure \ref{fig:fig3} presents a comparison between the best-performing nnUNet model (ensemble of Cascade and FullRes) and the worst-performing average ensemble of nnUNet and MedNeXt model trained on Dataset 506.

\begin{table}[h!]
\centering
\caption{$\text{DSC}_{\text{agg}}$ for nnUNet and MedNeXt trained on Dataset 506 for Task2}\label{tab6}
\begin{tabular}{|l|c|c|c|}
\hline
\textbf{Model} & \textbf{ GTVp $\text{DSC}_{\text{agg}}$ } & \textbf{ GTVn $\text{DSC}_{\text{agg}}$ } & \textbf{ Mean } \\
\hline
nnUNet FullRes & 0.6100 & 0.8508 & 0.7304 \\
nnUNet Cascade & 0.6105 & 0.8521 & 0.7313 \\
nnUNet ResEnc & 0.5742 & 0.8293 & 0.7018 \\
nnUNet All Ensembled & 0.6159 & 0.8544 & 0.7351 \\
nnUNet Cascade + FullRes Ensembled & \textbf{0.6173} & \textbf{0.8543} & \textbf{0.7358} \\
nnUNet Cascade + ResEnc Ensembled & 0.6031 & 0.8472 & 0.7251 \\
nnUNet FullRes + ResEnc Ensembled & 0.6022 & 0.8463 & 0.7242 \\
\hline
MedNeXt (Small with Kernel 3) & 0.6099 & 0.8306 & 0.7202 \\
\hline
nnUNet + MedNeXt (Average) & 0.6049 & 0.7735 & 0.6892 \\
\hline
\end{tabular}
\end{table}

\begin{table}[h!]
\centering
\caption{Mean DSC $\pm$ STD over all cases for nnUNet and MedNeXt trained on Dataset 506 for Task2}\label{tab8}
\begin{tabular}{|l|c|c|}
\hline
\textbf{Model} & \textbf{GTVp DSC} & \textbf{GTVn DSC} \\
\hline
nnUNet FullRes & 0.5148 $\pm$ 0.3374 & 0.8124 $\pm$ 0.1690 \\
nnUNet Cascade & 0.5134 $\pm$ 0.3324 & 0.8096 $\pm$ 0.1797 \\
nnUNet ResEnc & 0.4896 $\pm$ 0.3295 & 0.7869 $\pm$ 0.1886 \\
nnUNet All Ensembled & 0.5068 $\pm$ 0.3328 & \textbf{0.8167 $\pm$ 0.1619} \\
nnUNet Cascade + FullRes Ensembled & 0.5094 $\pm$ 0.3350 & 0.8139 $\pm$ 0.1719 \\
nnUNet Cascade + ResEnc Ensembled & 0.4992 $\pm$ 0.3342 & 0.8001 $\pm$ 0.1830 \\
nnUNet FullRes + ResEnc Ensembled & 0.4957 $\pm$ 0.3315 & 0.7996 $\pm$ 0.1834 \\
\hline
MedNeXt (Small with Kernel 3) & \textbf{0.5577} $\pm$ \textbf{0.3202} & 0.7804 $\pm$ 0.2150 \\
\hline
nnUNet + MedNeXt (Average) & 0.5147 $\pm$ 0.3525 & 0.7158 $\pm$ 0.3026 \\
\hline
\end{tabular}
\end{table}

For Task 2, the final submission used an nnUNet ensemble of FullRes and Cascade models. It achieved a $\text{DSC}_{\text{agg}}$ of \textbf{0.8519} for GTVn and \textbf{0.5491} for GTVp, with an overall mean $\text{DSC}_{\text{agg}}$ of \textbf{0.7005} in the final test phase on the 50 test patients.

\section{Discussion}

The results for Task 1 and Task 2 provide insights into the performance of various configurations of nnUNet and MedNeXt models, as well as the impact of pretraining with external datasets and ensembling strategies.

For \textbf{Task 1}, the MedNeXt small model with kernel size 3 achieved the best performance with the highest $\text{DSC}_{\text{agg}}$, outperforming all configurations of nnUNet, see Table \ref{tab1}. As a result, we chose the MedNeXt small kernel size 3 configuration as our final submission for Task 1.

Despite the fact that MedNeXt small model with kernel size 3 outperformed all nnUNet models, other MedNeXt architectures faced stability issues while training. Specifically, MedNeXt small model with kernel size 5, large model with kernel size 3, and large model with kernel size 5 repeatedly collapsed too early in training—after only a few epochs—so these models were not trained for enough epochs to be used effectively. Consequently, we were unable to fully compare the performance of different MedNeXt architectures and use the \texttt{mednextv1\_ensemble} method to aggregate predictions from various configurations of MedNeXt, as originally planned.

Interestingly, the average ensemble of nnUNet and MedNeXt predictions led to a lower $\text{DSC}_{\text{agg}}$ than using either model independently, suggesting that while both models have strengths, averaging their predictions may have introduced inconsistencies that reduced performance. Specifically, the average ensembling approach increased the number of false negatives and false positives, while also decreasing the number of true positives. This imbalance likely contributed to the overall drop in $\text{DSC}_{\text{agg}}$.

The comparison of mean DSC ± STD values shows that MedNeXt consistently achieved a higher mean DSC with less variability, for both GTVp and GTVn, see Table \ref{tab1.1}. This indicates greater robustness in its segmentation performance across different samples and superior overall performance. The nnUNet models had higher variability and lower mean DSC, particularly in GTVp predictions. In conclusion, MedNeXt proved to be a stronger candidate for reliable segmentation compared to nnUNet.

Pretrained models on the registered pre-RT and mid-RT images had higher $\text{DSC}_{\text{agg}}$ than those that were fine-tuned on original pre-RT images, see Table \ref{tab2}. 
This observation can be attributed to the fact that the pretraining process used all available input samples, without a separate validation set. Since the original pre-RT images and the registered pre-RT images are highly similar, it can be expected that the pretrained models, which were trained on the registered images, would perform better on a data that is quite similar to what was already seen. In other words, the evaluation on these pretrained models was not a reliable measure of their true performance.
In contrast, the fine-tuned models likely had lower $\text{DSC}_{\text{agg}}$ because they were fine-tuned using a 5-fold cross-validation setup. Finally, we concluded that fine-tuned models would most likely generalize better on unseen test data. 

When we experimented with pretraining nnUNet using the BraTS dataset, the results were mixed. Pretraining on BraTS alone led to poor performance, particularly for GTVn segmentation, see Table \ref{tab3}. Several factors likely contributed to this outcome. First, the BraTS dataset consists of images of the brain, while our challenge data includes the more anatomically complex H\&N region. Second, the BraTS dataset uses T1w MRI images, whereas our challenge data is T2w, which may have caused discrepancies in the features learned during pretraining. Finally, the BraTS dataset contains only one label (tumor region), whereas our challenge requires segmentation of two distinct labels (GTVp and GTVn). These differences likely inhibited the ability of the pretrained model to generalize well to the challenge-specific data. However, when BraTS was combined with mid-RT and pre-RT registered images, there was a notable improvement, although it still did not surpass the models trained solely on challenge-specific data. This dataset was less aligned with the challenge's specific needs, and thus its contribution to the final model performance was limited.


For \textbf{Task 2}, the impact of including registered pre-RT images and their segmentation masks in the multi-channel input was evaluated. It was observed that using registered pre-RT images alone, without their segmentation masks, did not provide useful information for segmenting mid-RT images. Using only the segmentation masks of registered pre-RT images along with the mid-RT images resulted in a significant improvement. However, including both registered pre-RT images and their segmentation masks further improved the performance of nnUNet FullRes, and it achieved the highest $\text{DSC}_{\text{agg}}$ for the mid-RT segmentation task when it was trained on dataset 506 (see Table \ref{tab5}). 

MedNeXt faced stability issues when trained on dataset 506, collapsing after a few hundred epochs for several folds, see Table \ref{tab4}.  Despite incomplete training, the MedNeXt model trained on Dataset 506 outperformed models trained on Datasets 505 and 504. Interestingly, the MedNeXt model trained on Dataset 507, achieved the best performance among MedNeXt models for Task 2. 
The observation that MedNeXt performed better on Dataset 507 than on Dataset 506, in contrast to nnUNet, is likely due to its inability to successfully complete 1000 epochs for all folds on Dataset 506. To address these training challenges, the dataset was refined by discarding samples with zero ground truth for either label, resulting in a stable training process for MedNeXt on Dataset 516. Nevertheless, MedNeXt showed its best performance when trained on Dataset 507.

Both nnUNet and MedNeXt models trained on datasets which included segmentation masks of registered pre-RT images (506 and 507) performed better than those trained on dataset 505 and 504. This further demonstrates the importance of including segmentation masks in the input data. This improvement can be attributed to the fact that the primary difference between mid-RT and pre-RT images lies in the size of the GTVn and GTVp, as mid-RT images are taken after some RT treatments. By incorporating pre-RT images and their segmentation, into the model’s input, the model can better understand the region of the GTVn and GTVp in the mid-RT images, and can be more accurate in localization and segmentation. 

Other MedNeXt architectures, specifically, MedNeXt small model with kernel size 5, large model with kernel size 3, and large model with kernel size 5 faced stability issues while training in Task 2 as well. They repeatedly collapsed too early in training—after only a few epochs—so these models were not trained
for enough epochs to be used effectively. As a result, it was not possible to compare the performance of different MedNeXt architectures and use the \texttt{mednextv1\_ensemble} method to aggregate predictions from various configurations of MedNeXt, as originally planned.

The comparison of different models highlights that nnUNet, particularly the nnUNet ensemble of FullRes and Cascade model, outperformed MedNeXt in terms of $\text{DSC}_{\text{agg}}$, which is the primary ranking metric for the challenge, see Table \ref{tab6}. As a result, the nnUNet ensemble of FullRes and Cascade was chosen as the final model for Task 2.

The ensemble of nnUNet and MedNeXt predictions for Task 2, using the average ensembling method, resulted in a lower $\text{DSC}_{\text{agg}}$ than using either model independently. This suggests that averaging predictions between models may not have fully captured their individual strengths and could have introduced inconsistencies in the final segmentation. Specifically, the average ensembling approach increased the number of false negatives and false positives, while reducing the number of true positives, ultimately lowering the $\text{DSC}_{\text{agg}}$ and overall performance.

The comparison of mean DSC ± STD values shows that nnUNet ensemble of all generally achieved a higher mean DSC with less variability, especially for GTVn, see Table \ref{tab8}. This indicates greater robustness in its segmentation performance across samples. For GTVp, MedNeXt outperformed nnUNet models. However, MedNeXt had higher variability and lower mean DSC, in GTVn predictions. This suggests that its stability issues and variability in results and lower $\text{DSC}_{\text{agg}}$ made it less reliable compared to the more robust and stable nnUNet ensembles. 


For practitioners looking to implement these approaches, the choice of model depends on the task requirements and available resources. MedNeXt is recommended for its robustness and high performance in GTVp and GTVn segmentation for Task 1. However, it is important to carefully monitor training stability when using larger configurations or datasets with imbalances. For Task 2, the nnUNet is suggested due to its consistent performance and ability to handle multi-channel inputs effectively.

The computational requirements for the two models are also different. MedNeXt needs more GPU memory and longer training time due to its complex design. Training a MedNeXt model takes about 180 seconds per epoch on an NVIDIA RTX 6000 GPU. In comparison, nnUNet takes about 60 seconds per epoch under the same conditions. nnUNet’s lower resource usage and faster training make it better for environments with limited resources, while MedNeXt is more suitable for tasks that require detailed spatial modeling and where sufficient resources are available.

\section{Conclusion}
In this study, we sought to address the issue of segmenting tumor volumes in HNC using MRI data for the purpose of adaptive RT planning. Two DL models, nnUNet and MedNeXt, were tested, with an investigation of diverse architectural configurations, ensemble methodologies, and pretraining on external dataset.

In conclusion, the nnUNet model, particularly when ensemble predictions are leveraged, demonstrated high efficacy in the segmentation tasks. MedNeXt also demonstrated potential, particularly in Task 1, but encountered challenges with stability during training for Task 2. Pretraining with domain-specific data proved to be a crucial step in Task 1, and the incorporation of registered pre-RT segmentation masks proved beneficial for enhancing the performance of both models for Task 2. This finding addresses the key aspect of this challenge, which explored whether incorporating prior time point data (pre-RT and mid-RT) into segmentation algorithms could enhance performance in RT applications. The results clearly show that using both registered pre-RT images and their segmentation masks significantly improves the model’s ability to accurately segment mid-RT images. Future work could focus on exploring more effective ensemble methods, such as weighted average ensemble, to better combine the strengths of both models. Additionally, addressing the stability issues faced by MedNeXt, particularly in Task 2, may involve adjusting the training process to prevent collapse during training. Improving the use of external datasets through more sophisticated domain adaptation could also enhance the effectiveness of pretraining.
\begin{credits}
\subsubsection{\ackname} 
This work is supported by the Plattform für KI-Translation Essen (KITE) from the REACT-EU initiative (EFRE-0801977, \url{https://kite.ikim.nrw/}) and ”NUM 2.0“ (FKZ: 01KX2121) and FWF enFaced 2.0 (grant number: KLI-1044, \url{https://enfaced2.ikim.nrw/}). André Ferreira thanks the Fundação para a Ciência e Tecnologia (FCT) Portugal for the grant 2022.11928.BD.

\end{credits}
%
%
%
\bibliographystyle{unsrt}
\bibliography{samplepaper}

\begin{thebibliography}{10}

\bibitem{Kiser2019DataFlood}
Kendall~J. Kiser, Benjamin~D. Smith, Jihong Wang, and Clifton~D. Fuller.
\newblock Après mois, le déluge: Preparing for the coming data flood in the mri-guided radiotherapy era.
\newblock {\em Frontiers in Oncology}, 9:983, 2019.

\bibitem{Pollard2017MRGuided}
Julianne~M. Pollard, Zhifei Wen, Ramaswamy Sadagopan, Jihong Wang, and Geoffrey~S. Ibbott.
\newblock The future of image-guided radiotherapy will be mr guided.
\newblock {\em British Journal of Radiology}, 90(1073):20160667, 2017.

\bibitem{Andrearczyk2022}
Vincent Andrearczyk, Valentin Oreiller, Moamen Abobakr, Azadeh Akhavanallaf, et~al.
\newblock Overview of the hecktor challenge at miccai 2022: Automatic head and neck tumor segmentation and outcome prediction in pet/ct.
\newblock In {\em Head and Neck Tumor Segmentation and Outcome Prediction. HECKTOR 2022. LNCS}, volume 13626, pages 1--30. Springer, Cham, 2023.

\bibitem{SegRap2023}
Xiangde Luo, Jia Fu, Yunxin Zhong, Shuolin Liu, Bing Han, Mehdi Astaraki, Simone Bendazzoli, Iuliana Toma-Dasu, et~al.
\newblock Segrap2023: A benchmark of organs-at-risk and gross tumor volume segmentation for radiotherapy planning of nasopharyngeal carcinoma.
\newblock In {\em MICCAI SegRap 2023}, 2023.

\bibitem{isensee2021nnunet}
Fabian Isensee, Paul~F. Jaeger, Simon~A. Kohl, Jens Petersen, and Klaus~H. Maier-Hein.
\newblock nnu-net: a self-configuring method for deep learning-based biomedical image segmentation.
\newblock {\em Nature methods}, 18(2):203--211, 2021.

\bibitem{Roy2023MedNeXt}
Subhrajit Roy, Georg Koehler, Christian Ulrich, Michael Baumgartner, Jens Petersen, Fabian Isensee, Paul~F. Jaeger, and Klaus~H. Maier-Hein.
\newblock Mednext: Transformer-driven scaling of convnets for medical image segmentation.
\newblock In {\em International Conference on Medical Image Computing and Computer-Assisted Intervention (MICCAI)}, 2023.

\bibitem{isensee2020nnunet}
Fabian Isensee, Paul~F. Jaeger, Simon~A. Kohl, Jens Petersen, and Klaus~H. Maier-Hein.
\newblock nnu-net: a self-configuring method for deep learning-based biomedical image segmentation.
\newblock {\em Nature Methods}, pages 1--9, 2020.

\bibitem{Li2020GenericEnsemble}
Ruizhe Li, Dorothee Auer, Christian Wagner, and Xin Chen.
\newblock A generic ensemble based deep convolutional neural network for semi-supervised medical image segmentation.
\newblock {\em arXiv preprint arXiv:2004.07995}, 2020.

\bibitem{Codella2019ISICChallenge}
Noel Codella, Veronica Rotemberg, Philipp Tschandl, M.~Emre Celebi, Stephen Dusza, David Gutman, Brian Helba, Aadi Kalloo, Konstantinos Liopyris, and Michael Marchetti.
\newblock Skin lesion analysis toward melanoma detection 2018: A challenge hosted by the international skin imaging collaboration (isic).
\newblock {\em arXiv preprint arXiv:1902.03368}, 2019.

\bibitem{Tschandl2018HAM10000}
Philipp Tschandl, Cliff Rosendahl, and Harald Kittler.
\newblock The ham10000 dataset, a large collection of multi-source dermatoscopic images of common pigmented skin lesions.
\newblock {\em Scientific Data}, 5:180161, 2018.

\bibitem{Astaraki2023SegRap}
Mehdi Astaraki, Simone Bendazzoli, and Iuliana Toma-Dasu.
\newblock Fully automatic segmentation of gross target volume and organs-at-risk for radiotherapy planning of nasopharyngeal carcinoma.
\newblock {\em arXiv:2310.02972}, 2023.

\bibitem{Myronenko2022HECKTOR}
Andriy Myronenko, Md~Mahfuzur~Rahman Siddiquee, Dong Yang, Yufan He, and Daguang Xu.
\newblock Automated head and neck tumor segmentation from 3d pet/ct: Hecktor 2022 challenge report.
\newblock {\em arXiv preprint arXiv:2209.10809}, 2022.

\bibitem{1398617}
Steve Pieper, Michael Halle, and Ron Kikinis.
\newblock 3d slicer.
\newblock In {\em 2004 2nd IEEE International Symposium on Biomedical Imaging: Nano to Macro (IEEE Cat No. 04EX821)}, pages 632--635 Vol. 1, 2004.

\bibitem{LaBella2024BraTS}
Dominic LaBella, Katherine Schumacher, Michael Mix, Kevin Leu, Shan McBurney-Lin, Pierre Nedelec, et~al.
\newblock Brain tumor segmentation (brats) challenge 2024: Meningioma radiotherapy planning automated segmentation.
\newblock {\em arXiv preprint arXiv:2405.18383}, 2024.

\bibitem{9871907}
Vincent Andrearczyk, Valentin Oreiller, Mario Jreige, Joël Castelli, John~O. Prior, and Adrien Depeursinge.
\newblock Segmentation and classification of head and neck nodal metastases and primary tumors in pet/ct.
\newblock In {\em 2022 44th Annual International Conference of the IEEE Engineering in Medicine \& Biology Society (EMBC)}, pages 4731--4735, 2022.

\bibitem{ed278621-dc3e-343f-ae66-540d8990b60d}
Lee~R. Dice.
\newblock Measures of the amount of ecologic association between species.
\newblock {\em Ecology}, 26(3):297--302, 1945.

\end{thebibliography}

\end{document}